\documentclass[11pt,a4paper]{article}

\usepackage[]{hyperref}
\usepackage{amsmath, amsthm, amssymb}
\usepackage{graphicx}
\usepackage{setspace}

\def\ket#1{\left|#1\right>}
\def\bra#1{\langle#1\vert}

\def\expt(#1#2){\langle #2 \vert #1 \vert #2 \rangle}

\title{Can the ontology of Bohmian mechanics consists only in particles? The PBR theorem says no}

\author{Shan Gao
\\Research Center for Philosophy of Science and Technology, 
\\ Shanxi University, Taiyuan 030006, P. R. China
\\ E-mail:  \href{mailto:gaoshan2017@sxu.edu.cn}{gaoshan2017@sxu.edu.cn}.}

\begin{document}

 
\maketitle

\begin{abstract}\noindent 

The meaning of the wave function is an important unresolved issue in Bohmian mechanics. On the one hand, according to the nomological view, the wave function of the universe or the universal wave function is nomological, like a law of nature. On the other hand, the PBR theorem proves that the wave function in quantum mechanics or the effective wave function in Bohmian mechanics is ontic, representing the ontic state of a physical system in the universe. It is usually thought that the nomological view of the universal wave function is compatible with the ontic view of the effective wave function, and thus the PBR theorem has no implications for the nomological view. In this paper, I argue that this is not the case, and these two views are in fact incompatible. This means that if the effective wave function is ontic as the PBR theorem proves, then the universal wave function cannot be nomological, and the ontology of Bohmian mechanics cannot consist only in particles. This incompatibility result holds true not only for Humeanism and dispositionalism but also for primitivism about laws of nature, which attributes a fundamental ontic role to the universal wave function. Moreover, I argue that although the nomological view can be held by rejecting one key assumption of the PBR theorem, the rejection will lead to serious problems, such as that the results of measurements and their probabilities cannot be explained in ontology in Bohmian mechanics. Finally, I briefly discuss three $\psi$-ontologies, namely a physical field in a fundamental high-dimensional space, a multi-field in three-dimensional space, and RDMP (Random Discontinuous Motion of Particles) in three-dimensional space, and argue that the RDMP ontology can answer the objections to the $\psi$-ontology raised by the proponents of the nomological view. 

\end{abstract}

\vspace{6mm}

\section{Introduction}

Bohmian mechanics, being a modern formulation of the pilot-wave theory of de Broglie and Bohm (de Broglie, 1928; Bohm, 1952), provides an ontology of particles and their trajectories in space and time for  quantum mechanics (D\"{u}rr, Goldstein and Zangh\`{i}, 1992; Goldstein, 2021).  
One important unresolved issue in this theory is the meaning of the wave function. 
Is the wave function of the universe ontic, representing a concrete physical entity, or nomological, like a law of nature?\footnote{Throughout this paper, the word ``ontic'' or ``ontology'' denotes only material ontology which does not contain laws of nature unless otherwise stated. This is consistent with the use of the word in the PBR theorem and relevant literature. Note that some Bohmians use the word ``ontology'' to denote a general ontology which contains also laws of nature (Maudlin, 2007; Allori et al, 2008).}  
In recent years, the nomological view of the wave function becomes more and more popular (Allori et al, 2008; Esfeld et al, 2014). On this view, 
Bohmian mechanics is committed only to particles' positions and a law of motion. 
At the same time, a general and rigorous approach called ontological models framework has been proposed to determine the relation between the wave function and the ontic state of a physical system (Harrigan and Spekkens, 2010), and several $\psi$-ontology theorems have been proved in the framework (Pusey, Barrett and Rudolph, 2012; Colbeck and Renner, 2012, 2017; Hardy, 2013). In particular, the Pusey-Barrett-Rudolph theorem or the PBR theorem proves that the wave function in quantum mechanics or the effective wave function in Bohmian mechanics is ontic, representing the ontic state of a physical system in the universe (Pusey, Barrett and Rudolph, 2012). 
An interesting question then arises: is the nomological view of the universal wave function is compatible with the ontic view of the effective wave function? 

This issue has not received much attention in the literature. 
Goldstein's (2021) comprehensive review of Bohmian mechanics does not mention the PBR theorem. 
Esfeld et al's (2014) insightful paper about the ontology of Bohmian mechanics refers to the theorem once but without discussion.  
Callender (2015) gave an interesting but very brief analysis of the relevance of the PBR theorem to Bohmian mechanics. 
Presumably it is thought that the two views of the wave function are obviously compatible. 
According to Callender (2015), the difference between the ontic view and the nomological view of the wave function lies only in the metaphysics of these views; although the nomological view denies that the wave function exists as a concrete physical entity, it agrees that ``the wavefunction is part of the ontic state''.  
Moreover, according to Esfeld et al (2014), although Bohmian mechanics says that the universal wave function is nomological, it regards the effective wave function of a subsystem in the universe as ontic, representing ``an objective, physical degree of freedom belonging to the subsystem'', and thus the nomological view is compatible with the ontic view. 
In this paper, I will argue that this received view is debatable, and a careful analysis of the (in)compatibility between the nomological view and the ontic view will deepen our understandings of the meaning of the wave function in Bohmian mechanics. 

The rest of this paper is organized as follows. 
In Section 2, I first introduce Bohmian mechanics and the nomological view of the wave function. According to the nomological view, the universal wave function is nomological, like a law of nature, and the ontology of Bohmian mechanics consists only in particles. In Section 3, I then introduce the ontological models framework and the PBR theorem. The theorem proves that the effective wave function in Bohmian mechanics is ontic, representing the ontic state of a physical system in the universe. 
In Section 4, I argue that the nomological view of the universal wave function is incompatible with the ontic view of the effective wave functions. 
In Section 5, I argue that the incompatibility result holds true not only for Humeanism and dispositionalism but also for primitivism about laws of nature, which attributes a fundamental ontic role to the universal wave function. 
In Section 6, I point out that the nomological view can be held by rejecting one key assumption of the PBR theorem. But the rejection will arguably lead to serious problems, such as that the results of measurements and their probabilities cannot be explained in ontology in Bohmian mechanics. 
In Section 7, I briefly discuss three $\psi$-ontologies, namely a physical field in a fundamental high-dimensional space, a multi-field in three-dimensional space, and RDMP (Random Discontinuous Motion of Particles) in three-dimensional space, and argue that the RDMP ontology can answer the objections to the $\psi$-ontology raised by the proponents of the nomological view. 
Conclusions are given in the last section. 

\section{Bohmian mechanics and the nomological view}

In Bohmian mechanics, a complete realistic description of a quantum system is provided by the configuration defined by the positions of its particles together with its wave function. The law of motion is expressed by two equations: a guiding equation for the configuration of particles and the Schr\"{o}dinger equation, describing the time evolution of the wave function which enters the guiding equation. The law of motion can be formulated as follows:
              
\begin{equation}
{{dQ(t)} \over {dt}}=v^{\Psi(t)}(Q(t)),
\label{Ge}
\end{equation}

\begin{equation}
i\hbar {\partial \Psi(t) \over \partial t}=H\Psi(t),
\label{Sc}
\end{equation}

\noindent where $Q(t)$ denotes the spatial configuration of particles, $\Psi(t)$ is the wave function of the particle configuration at time $t$, and $v$ equals to the velocity of probability density in quantum mechanics. 
Moreover, it is postulated that at some initial instant $t_0$, the epistemic probability of the configuration, $\rho(t_0)$, is given by the Born rule: $\rho(t_0)=|\Psi(t_0)|^2$. This is called quantum equilibrium hypothesis, which, together with the law of motion, ensures the empirical equivalence between Bohmian mechanics and quantum mechanics. 

The status of the above equations is different, depending on whether one considers the physical description of the universe as a whole or of a subsystem thereof. Bohmian mechanics starts from the concept of a universal wave function or the wave function of the universe, figuring in the fundamental law of motion for all the particles in the universe. That is, $Q(t)$ describes the configuration of all the particles in the universe at time $t$, and $\Psi(t)$ is the wave function of the universe at time $t$, guiding the motion of all particles taken together. In this case, the Schr\"{o}dinger equation will be replaced by another equation in quantum cosmology such as the Wheeler-DeWitt equation. 
To describe subsystems of the universe, the appropriate concept is the effective wave function in Bohmian mechanics. 

The effective wave function is the Bohmian analogue of the usual wave function in quantum mechanics. It is not primitive, but derived from the universal wave function and the actual spatial configuration of all the particles ignored in the description of the respective subsystem (D\"{u}rr, Goldstein and Zangh\`{i}, 1992). The effective wave function of a subsystem can be defined as follows. Let $A$  be a subsystem of the universe including $N$ particles with position variables $x=(x_1,x_2,...,x_N)$. Let $y=(y_1,y_2,...,y_M)$ be the position variables of all other particles not belonging to $A$. Then the subsystem $A$'s conditional wave function at time $t$ is defined as the universal wave function $\Psi_t(x, y)$ evaluated at $y = Y(t)$:

\begin{equation}
\psi_t^A(x)=\Psi_t(x, y)|_{y=Y(t)}.
\end{equation}

\noindent If the universal wave function can be decomposed in the following form:

\begin{equation}\label{EF}
\Psi_t(x, y)=\varphi_t(x)\phi_t(y)+\Theta_t(x, y),
\end{equation}

\noindent where $\phi_t(y)$ and $\Theta_t(x, y)$ are functions with macroscopically disjoint supports, and $Y(t)$ lies within the support of $\phi_t(y)$, then $\psi_t^A(x)=\varphi_t(x)$ (up to a multiplicative constant) is $A$'s effective wave function at $t$. It can be seen that the temporal evolution of $A$'s particles is given in terms of $A$'s conditional wave function in the usual Bohmian way, and when the conditional wave function is  $A$'s effective wave function,  it also obeys a Schr\"{o}dinger dynamics of its own. This means that the effective descriptions of subsystems are of the same form of the law of motion as given above. 

Bohmian mechanics raises the question of the status of the wave function that figures in the law. 
According to the nomological view of the wave function, the relationship between the universal wave function and the motion of the particles should be conceived as a nomic one, instead of a causal one in terms of one physical entity acting on the other (D\"{u}rr, Goldstein and Zangh\`{i}, 1997; Goldstein and Teufel, 2001; Goldstein and Zangh\`{i}, 2013; Esfeld et al, 2014). 
In the words of D\"{u}rr, Goldstein and Zangh\`{i} (1997),

\begin{quote}
The wave function of the universe is not an element of physical reality. We propose that the wave function belongs to an altogether different category of existence than that of substantive physical entities, and that its existence is nomological rather than material. We propose, in other words, that the wave function is a component of a physical law rather than of the reality described by the law. (p. 10)
\end{quote}

As argued by these authors, the reasons to adopt this nomological view of the wave function come from the unusual kind of way in which Bohmian mechanics is formulated, and the unusual kind of behavior that the wave function undergoes in the theory (D\"{u}rr, Goldstein and Zangh\`{i}, 1997). 
First, although the wave function affects the behavior of the configuration of the particles, which is expressed by the guiding equation (\ref{Ge}), there is no back action of the configuration upon the wave function. The evolution of the wave function is governed by the Schr\"{o}dinger equation (\ref{Sc}), in which the actual configuration $Q(t)$ does not appear. Since a physical entity is supposed to satisfy the action-reaction principle, it seems that the wave function cannot describe a physical entity in Bohmian mechanics. Next, the wave function of a many-particle system, $\psi(q_1, . . . , q_N)$, is defined not in our ordinary three-dimensional space, but in the $3N$-dimensional configuration space, the set of all hypothetical configurations of the system. Thus it seems untenable to view the wave function as directly describing a real physical field. 

Thirdly, the wave function in Bohmian mechanics plays a role that is analogous to that of the Hamiltonian in classical Hamiltonian mechanics (Goldstein\index{Goldstein, Sheldon} and Zangh\`{i}\index{Zangh\`{i}, Nino}, 2013). Both the classical Hamiltonian and the wave function live on a high dimensional space. The wave function is defined in configuration space, while the classical Hamiltonian is defined in phase space: a space that has twice as many dimensions as configuration space. 
Besides, there is a striking analogy between the guiding equation in Bohmian mechanics and the Hamiltonian equations in classical mechanics. 
Since the classical Hamiltonian is regarded not as a description of some physical entity, but as the generator of time evolution in classical mechanics, it seems natural to assume that the wave function is not a description of some physical entity either, but a similar generator of the equations of motion in Bohmian mechanics. 

Admittedly, these arguments for the nomological view of the wave function are not conclusive. 
In fact, they have been criticized by a few authors. 
For example, Romano (2021) posed an objection to the above analogy between the wave
function in Bohmian mechanics and the Hamiltonian function in classical mechanics. 
According to his analysis, 
the classical Hamiltonian, being the sum of the kinetic and potential energy, is just a mathematically useful way to rewrite the physical properties of the particles composing the classical system, and it can be in principle eliminated and substituted by a direct description of particles’ properties and forces acting on them. 
By contrast, the wave function is derived as a solution of the Schr\"{o}dinger equation, and it does not refer to the properties of the Bohmian particles and cannot be eliminated by the theory either. 
Thus, although the classical Hamiltonian is arguably a nomological entity, it is not clear why the wave function should be also regarded as a nomological entity. 

In addition, several authors have also argued against the nomological view of the wave function based on an analysis of the properties of the universal wave function (Hubert and Romano, 2018; Valentini, 2020). 
According to Goldstein and Zangh\`{i} (2013), 
the wave function of the universe is not controllable, and it may be not dynamical either. 
They illustrated the latter point by the Wheeler-DeWitt equation, which is the fundamental equation for the wave function of the universe in canonical quantum cosmology:

\begin{equation}
H\Psi(q)=0,
\label{WD}
\end{equation}

\noindent where $\Psi(q)$ is the wave function of the universe, $q$ refers to 3-geometries, and $H$ is the Hamiltonian constraint which involves no explicit time-dependence. Unlike the Schr\"{o}dinger equation, the Wheeler-DeWitt equation has on one side, instead of a time derivative of $\Psi$, simply 0, and thus its natural solutions are time-independent. 
Moreover, the wave function of the universe may be unique. Although the Wheeler-DeWitt equation presumably has a great many solutions, when supplemented with additional natural conditions such as the Hartle-Hawking boundary condition, the solution may become unique. Such uniqueness also fits with the conception of the universal wave function as law. 

However, Valentini (2020) argued that the universal wave function is contingent, and it is neither time-independent nor uniquely determined by the laws of quantum gravity. 
He noted that many researchers had suggested that a physical time parameter is in effect
hidden within the quantum-gravitational formalism and that the wave function is in fact time-dependent when correctly written as a function of physical degrees of freedom. 
He also emphasized that the key aspect of the universal wave function that makes it count as a physical entity is its contingency or its under-determination by known physical laws. 
Besides, as pointed out by Hubert and Romano (2018), even if the universal wave function is time-independent, it is still a solution of a dynamical equation such as the Wheeler-DeWitt equation, while having a law for a nomological entity seems to be not in the spirit of a nomological entity. This is different for the Hamiltonian in classical mechanics, which does not arise from a law, and therefore can be interpreted as a nomological entity. 

Notwithstanding these objections, 
the nomological view of the wave function is still a popular view. 
According to this view, the universal wave function is nomological, describing a law and not describing some sort of concrete physical entity in Bohmian mechanics. 
A law of motion tells us what happens in space and time given the specification of initial conditions, but it is not itself a physical entity existing in space and time. 
The exact meaning of the universal wave function then depends on what exactly a law is. 
There are two main views about laws of nature in the literature, namely Humeanism and dispositionalism, and both of them can be drawn upon for developing the nomological view of the wave function in Bohmian mechanics (Esfeld et al, 2014). By Humeanism about laws, there are only particles' positions in the ontology, while dispositionalism admits more in the ontology than particles' positions, namely the holistic disposition of all the particles in the universe. 
My following analysis of Bohmian mechanics and the nomological view of the wave function applies to these two views about laws of nature.\footnote{Note that Bohmian mechanics is also compatible with primitivism about laws as suggested by Maudlin (2007). I will discuss whether my analysis applies to this view additionally.}  
  
\section{The PBR theorem}

Although there are various reasons to adopt the nomological view of the universal wave function in Bohmian mechanics, we in fact know little about the universal wave function itself, since the final theory of quantum gravity is not yet available. 
A more feasible approach is to analyze the meaning of the effective wave function of a subsystem in the universe, such as whether the effective wave function has a tenable physical explanation under the nomological view of the universal wave function.\footnote{Romano (2021) already noticed that the metaphysical status of effective wave functions in Bohmian mechanics is not well defined within the nomological view of the universal wave function, although he did not give an analysis of the issue.}   
In recent years, there appear several rigorous arguments supporting the ontic view of the wave function in quantum mechanics or the effective wave function in Bohmian mechanics, which has been called $\psi$-ontology theorems. 
In this section, I will introduce the ontological models framework and an important $\psi$-ontology theorem proved in the framework, the PBR theorem.\footnote{Here it is worth pointing out that several authors have recently argued that the ontological models framework and the PBR theorem have certain limitations, and in particular, they do not apply to Rovelli's (1996) relational quantum mechanics, which employs ontic states dealing with relational properties but regards the wave function as a computational device encoding observers' information (Oldofredi and L\'{o}pez, 2020; Oldofredi and Calosi, 2021). However, these limitations do not affect my analysis of the implications of the PBR theorem for the nomological view of the wave function.}  

Quantum mechanics, in its minimum formulation, is an algorithm for calculating probabilities of measurement results. 
The theory assigns a mathematical object, the wave function, to a physical system appropriately prepared at a given instant, and specifies how the wave function evolves with time. The time evolution of the wave function is governed by the Schr\"{o}dinger equation, and the connection of the wave function with the results of measurements on the system is specified by the Born rule. 
At first sight, quantum mechanics as an algorithm says nothing about the actual ontic state of a physical system. 
However, it has been known that this is not true due to the recent advances in the research of the foundations of quantum mechanics (see Leifer, 2014 for a helpful review). 

First of all, a general and rigorous approach called ontological models framework has been proposed to determine the relation between the wave function and the ontic state of a physical system (Harrigan and Spekkens 2010). 
The framework has two fundamental assumptions. 
The first assumption is about the existence of the underlying state of reality. It says that if a physical system is prepared such that quantum mechanics assigns a wave function to it, then after preparation the system has a well-defined set of physical properties or an underlying ontic state, which is usually represented by a mathematical object, $\lambda$. 
In general, for an ensemble of identically prepared systems to which the same wave function $\psi$ is assigned, the ontic states of different systems in the ensemble may be different, and the wave function $\psi$ corresponds to a probability distribution $p(\lambda|\psi)$ over all possible ontic states, where $\int{d\lambda p(\lambda|\psi)}=1$. 

There are two possible types of models in the ontological models framework, namely $\psi$-ontic models and $\psi$-epistemic models. 
In a $\psi$-ontic model, the ontic state of a physical system uniquely determines its wave function, and the probability distributions corresponding to two different wave functions do not overlap. 
In this case, the wave function directly represents the ontic state of the system. While in a $\psi$-epistemic model, the probability distributions corresponding to two different wave functions may overlap, and there are at least two wave functions which are compatible with the same ontic state of a physical system. 
In this case, the wave function merely represents a state of incomplete knowledge - an epistemic state - about the actual ontic state of the system. 

In order to investigate whether an ontological model is consistent with the predictions of quantum mechanics, we also need a rule of connecting the underlying ontic states with measurement results. 
This is the second assumption of the ontological models framework, which says that when a measurement is performed, the behaviour of the measuring device is determined by the ontic state of the system, along with the physical properties of the measuring device. 
Concretely speaking, for a projective measurement $M$, the ontic state $\lambda$ of a physical system determines the probability $p(k|\lambda,M)$ of different results $k$ for the measurement $M$ on the system. 
The consistency with quantum mechanics then requires the following relation: 

\begin{equation}
\int{d\lambda p(k|\lambda, M)p(\lambda|\psi)} = p(k|M,\psi),
\label{CD}
\end{equation}

\noindent where $p(k|M,\psi)=|\bra{k}\psi\rangle|^2$ is the Born probability of $k$ given $M$ and the wave function $\psi$. 

Second, several important $\psi$-ontology theorems have been proved in the ontological models framework (Pusey, Barrett and Rudolph, 2012; Colbeck and Renner, 2012, 2017; Hardy, 2013), one of which is the PBR theorem (Pusey, Barrett and Rudolph, 2012).  
The PBR theorem shows that in the ontological models framework, when assuming independently prepared systems have independent ontic states, the ontic state of a physical system uniquely determines its wave function, or the wave function of a physical system directly represents the ontic state of the system. 
This auxiliary assumption is called preparation independence assumption. 

The basic proof strategy of the PBR theorem is as follows. Assume there are $N$ nonorthogonal quantum states $\psi_i$ ($i$=1, ... , $N$), which are compatible with the same ontic state $\lambda$.\footnote{It can be readily shown that different orthogonal states correspond to different ontic states based on the  ontological models framework. Thus the proof given here concerns only nonorthogonal states.} The ontic state $\lambda$ determines the probability $p(k|\lambda,M)$ of different results $k$ for the measurement $M$. Moreover, there is a normalization relation for any $N$ result measurement: $\sum_{i=1}^{N}p(k_i|\lambda,M)=1$. Now if an $N$ result measurement satisfies the condition that the first state gives zero Born probability to the first result and the second state gives zero Born probability to the second result and so on, then there will be a relation $p(k_i|\lambda,M)=0$ for any $i$, which leads to a contradiction.

The task is to find whether there are such nonorthogonal states and the corresponding measurement. Obviously there is no such a measurement for two nonorthogonal states of a physical system, since this will permit them to be perfectly distinguished, which is prohibited by quantum mechanics. However, such a measurement does exist for four nonorthogonal states of two copies of a physical system. The four nonorthogonal states are the following product states: $\ket{0} \otimes \ket{0}$, $\ket{0} \otimes \ket{+}$,$\ket{+} \otimes \ket{0}$ and $\ket{+} \otimes \ket{+}$, where $\ket{+}= {1 \over \sqrt{2}}(\ket{0} + \ket{1})$. The corresponding measurement is a joint measurement of the two systems, which projects onto the following four orthogonal states:

\begin{eqnarray}\label{}
\phi_1& =& \tfrac{1}{\sqrt{2}}(\ket{0}\otimes\ket{1}+\ket{1}\otimes\ket{0}), \nonumber\\
\phi_2& =& \tfrac{1}{\sqrt{2}}(\ket{0}\otimes\ket{-}+\ket{1}\otimes\ket{+}), \nonumber\\
\phi_3& =& \tfrac{1}{\sqrt{2}}(\ket{+}\otimes\ket{1}+\ket{-}\otimes\ket{0}), \nonumber\\
\phi_4& =& \tfrac{1}{\sqrt{2}}(\ket{+}\otimes\ket{-}+\ket{-}\otimes\ket{+}),
\end{eqnarray}

\noindent where $\ket{-}={1 \over \sqrt{2}}(\ket{0}-\ket{1})$. This proves that the four nonorthogonal states are ontologically distinct. In order to further prove the two nonorthogonal states $\ket{0}$ and $\ket{+}$ for one system are ontologically distinct, the preparation independence assumption is needed. Under this assumption, a similar proof for every pair of nonorthogonal states can also be found, which requires more than two copies of a physical system (see Pusey, Barrett and Rudolph, 2012 for the complete proof). 

To sum up, the PBR theorem shows that quantum mechanics as an algorithm may also say something about the ontic state of a physical system. It is that under the preparation independence assumption, the wave function assigned to a physical system, which is used for calculating probabilities of results of measurements on the system, is a mathematical representation of the ontic state of the system in the ontological models framework. 

\section{Is the nomological view compatible with the ontic view?}

The PBR theorem raises an intriguing question for Bohmian mechanics: 
if the effective wave function is ontic as the PBR theorem proves, can the universal wave function be nomological? or is the nomological view of the universal wave function compatible with the ontic view of the effective wave function? 
The received view seems to be that these two views are obviously compatible, and thus the PBR theorem has no implications for the nomological view of the universal wave function in Bohmian mechanics. 
In the following, I will argue that this is not the case. 

First of all, it is worth emphasizing that the PBR theorem does not only reject the $\psi$-epistemic view, but also \emph{establish} the $\psi$-ontic view (given its assumptions). In particular, it says that the wave function is ontic in the sense of material ontology, not in the sense of a general ontology which also contains laws of nature. This can be seen from one key assumption of the ontological models framework on which the PBR theorem is based, which says that the ontic state of a physical system determines the results of measurements on the system and their probabilities. By comparison, a nomological fact or ontology, although it is real, has no such specific causal efficacy of a concrete physical entity. 
In other words, the response of a measuring device to a physical system should be determined by the  ontic state of \emph{this} system, while a law of nature only describes or dictates \emph{how} the measuring device responds to the system given their initial ontic states, and it does not \emph{cause} the specific response of the device.  

Next, it can be seen that if the PBR theorem applies not only to the subsystems of the universe, but also to the universe as a whole, then it is obvious that the ontic view and the nomological view are two different views of the universal wave function, and thus they are incompatible. 
According to the ontic view, the universal wave function is ontic, representing the ontic state of a concrete physical system, the universe, while according to the nomological view, the universal wave function is  nomological, and it does not represent the ontic state of the universe. 
However, it is arguable that the PBR theorem may not apply to the universe as a whole, since the proof of the theorem concerns measurements and copies of a single physical system, while there is only one universe and no measurements can be made on the universe as a whole either. 

Now consider the effective wave function of a subsystem in the universe. 
In order that the nomological view of the universal wave function is compatible with the ontic view of the effective wave function, the effective wave function of a subsystem must represent a physical property of  Bohmian particles, since the ontology of Bohmian mechanics consists only in particles on the nomological view.\footnote{If the properties of Bohmian particles are only position and velocity as usually thought, then it will be obvious that the effective wave function does not represent a physical property of these particles, and thus the nomological view of the universal wave function is not compatible with the ontic view of the effective wave function. But, as Esfeld et al (2014) thought, there may exist other properties of Bohmian particles besides position and velocity, such as nonlocal interactions between these particles, so that the effective wave function of a subsystem encodes the nonlocal influence of other particles on the subsystem. The purpose of my following analysis is to exclude this possibility.} 
If the effective wave function of a subsystem represents a physical property of another physical entity different from particles, then the nomological view of the universal wave function, which requires that there are only particles in the ontology of Bohmian mechanics, cannot be true. 

Let's see if the effective wave function of a subsystem represents a physical property of Bohmian particles.  
First, as Esfeld et al (2014) have already pointed out, 
the effective wave function of a subsystem is clearly not internal degrees of freedom of the particles of the subsystem. 
Indeed, since the effective wave function of a subsystem depends on the universal wave function and the configuration of all other particles in the universe,\footnote{In this sense, it is sometimes said that the effective wave function is quasi-nomological. However, as noted by Romano (2021), it is still not clear what ``quasi-nomological'' means. This paper does not aim to clarify the quasi-nomological status 
of effective wave functions but aims to argue that no matter what ``quasi-nomological'' means, the nomological view is inconsistent with the result of the PBR theorem.} it is impossible to interpret it as a representation of certain intrinsic property of the particles of the subsystem. 

Then, the only possibility is that the effective wave function of a subsystem represents a physical property of the particles in the environment of the subsystem (i.e. other particles in the universe).   
Since the effective wave function of a subsystem influences the motion of the particles of the subsystem by the guiding equation, this physical property of the particles in the environment must have the efficacy of influencing the particles of the subsystem. 
Moreover, since the particles in the environment and the particles of the subsystem are spacelike separated at each instant, the influence, if it exists, must be nonlocal. 

This is the view supported by Esfeld et al (2014). 
According to these authors, the effective wave function of a subsystem encodes the nonlocal influence of other particles on the subsystem via the nonlocal law of Bohmian mechanics. 
For example, in the double-slit experiment with one particle at a time, the particle goes through exactly one of the two slits, and that is all there is in the physical world. There is no real physical field that guides the motion of the particle and propagates through both slits and undergoes interference. The development of the position of the particle (its velocity and thus its trajectory) is determined by the positions of other particles in the universe, including the particles composing the experimental setup, and the nonlocal law of Bohmian mechanics can account for the observed particle position on the screen (Esfeld et al, 2014). 
If this view is true, then one can say that the nomological view of the universal wave function regards the effective wave function as ontic, and thus it is consistent with the ontic view of the effective wave function. 

However, it can be argued that the effective wave function of a subsystem does not represent a physical property of the particles in the environment, such as encoding the nonlocal influence of these particles on the subsystem.   
Let's first consider the simplest case in which the universal wave function factorizes so that

\begin{equation}
\Psi_t(x, y)=\varphi_t(x)\phi_t(y),
\end{equation}

\noindent where $x=(x_1,x_2,...,x_N)$ is the position variables of $N$ particles of a subsystem $A$ of the universe, and $y=(y_1,y_2,...,y_M)$ is the position variables of all other particles not belonging to $A$. 
Then $\psi_t^A(x)=\varphi_t(x)$ is subsystem $A$'s effective wave function at $t$.  In this case, subsystem $A$ and its environment, which are in a product state, are independent of each other. Thus, the effective wave function of subsystem $A$ is independent of the particles in the environment, and it does not represent a physical property of these particles. 
Moreover, since subsystem $A$ and its environment are in a product state, the particles in the environment do not have nonlocal influence on the particles of subsystem $A$, and thus the effective wave function of subsystem $A$ cannot encode such a non-existent nonlocal influence either.  

Next, consider the general case in which there is an extra term in the factorization of the universal wave function:

 \begin{equation}
\Psi_t(x, y)=\varphi_t(x)\phi_t(y)+\Theta_t(x, y). 
\end{equation}

\noindent In this case, the effective wave function of subsystem A is determined by both the universal wave function $\Psi_t(x, y)$ and the positions of the particles in its environment $Y(t)$. If $Y(t)$ lies within the support of $\phi_t(y)$, $A$'s effective wave function at $t$ will be $\varphi_t(x)$. If $Y(t)$ does not lie within the support of $\phi_t(y)$, $A$'s effective wave function at $t$ will be not $\varphi_t(x)$. For example, suppose $\Theta_t(x, y)=\sum_n{f_n(x)g_n(y)}$, where $g_i(y)$ and $g_j(y)$ are functions with macroscopically disjoint supports for any $i \neq j$, then if $Y(t)$ lies within the support of $g_i(y)$, $A$'s effective wave function at $t$ will be $f_i(x)$. 

It can be seen that the role played by the particles in the environment is only selecting which function the effective wave function of subsystem A is, while each selected function is completely determined by the universal wave function. 
Thus the effective wave function of subsystem $A$ as part of the universal wave function will only represent a physical property of the part of something represented by the universal wave function (if there is any), and it does not represent a physical property of the particles in the environment. 
This is like the parable of blind men touching an elephant. 
If a blind man touches the tail of an elephant, he would say that the elephant is like a rope. The property of being like a rope is a property of one part of the elephant, and it is not the property of the blind man. 

Moreover, even if the effective wave function of subsystem $A$ represents a physical property of the particles in the environment, this property has no efficacy of influencing the particles of the subsystem nonlocally. 
According to the Bohmian law of motion, when the effective wave function of subsystem $A$ has been selected (via a measurement-like process), the particles in the environment have no nonlocal influence on the particles of subsystem $A$; the particles of the subsystem and the particles in the environment reside in an effective product state such as $\varphi_t(x)\phi_t(y)$. 
For example, in the double-slit experiment with one particle at a time, the development of the position of the particle will not depend on the positions of other particles in the universe (if only the positions of these particles select the same effective wave function of the particle during the experiment, e.g. $Y(t)$ has been within the support of $\phi_t(y)$ during the experiment). 

Finally, it is worth noting that 
even if the effective wave function of subsystem $A$ encodes the nonlocal influence of the particles in the environment on the subsystem, it does not imply that the whole effective wave function is a property of these particles. 
The reason is that the nonlocal influence, which determines the velocities of the particles of the subsystem, is determined only by the phase of the effective wave function, and not by the amplitude of the effective wave function. 
Thus, even if there were such a nonlocal influence, it would only indicate that the phase of the effective wave function of subsystem $A$ represents a property of the particles in the environment, and it does not imply that the amplitude of the effective wave function of subsystem $A$ also represents a property of these particles. 

To sum up, I have argued that 
the effective wave function of a subsystem represents neither a physical property of 
the particles of the subsystem nor a physical property of the particles in the environment, such as encoding their nonlocal influence on the subsystem. 
In short, the effective wave function of a subsystem does not represent a physical property of Bohmian particles. 
This means that the nomological view of the universal wave function is not compatible with the ontic view of the effective wave function. 
If the effective wave function is ontic as the PBR theorem proves, then 
it must represent a property of another physical entity different from Bohmian particles, and thus the universal wave function cannot be nomological, and the ontology of Bohmian mechanics cannot consist only in particles. 

\section{On primitivism} 

Although the above analysis of the nomological view of the universal wave function applies to the two main views about laws of nature, namely Humeanism and dispositionalism, we need a further analysis of whether it also applies to primitivism (Maudlin, 2007), which, unlike Humeanism and dispositionalism, attributes a fundamental ontic role to the universal wave function. 
According to primitivism, laws of nature are irreducible global nomic facts in each physically possible world, which are ontologically prior with respect to local particular facts such as an initial configuration of particles in a background spacetime (Dorato and Esfeld, 2015). 
The key is to determine what ontology the universal wave function corresponds to on primitivism. 

It has been argued that primitivism commits us to the view that laws of nature, including the universal wave function viewed as a nomological thing, are mathematical entities, not physical entities (Dorato, 2015).  
Here an entity is physical if and only if it is either in spacetime or causally active or both, and an entity is mathematical if and only if it is not physical.
If this argument is valid, then although the universal wave function exists objectively in the world on primitivism, it is neither in spacetime nor causally active, and thus it should not be included in the ontic states of all subsystems of the universe, which are required to be causally active such as shifting the pointer of a measuring device.
In other words, the ontology of Bohmian mechanics still consists only in particles on primitivism. Then, the previous analysis will be also valid for primitivism, as for Humeanism and dispositionalism. 

Furthermore, it can be seen that 
assuming the universal wave function has causal influences on the particles in spacetime and thus is part of the ontic state is against the original motivation of the nomological view of the universal wave function. 
As noted before, there are three main reasons to avoid the $\psi$-ontology and assume the nomological view. 
First, the wave function of $N$ particles is defined not in three-dimensional space, but in the $3N$-dimensional configuration space. It is thus unclear how a physical field in a $3N$-dimensional space can affect the behavior of the particles in three-dimensional space. Second, even if a physical field can affect the behavior of the particles, there is no back action of the particles upon the physical field, and this violates the action-reaction principle. Third, the wave function plays a role that is analogous to that of the classical Hamiltonian, which is regarded not as a description of some physical entity in classical mechanics. 

Now, if assuming that the universal wave function has causal influences on the particles, then the above objections to the $\psi$-ontology will be also objections to primitivism with this assumption. First, it is unclear how a law of nature not existing in space can have causal influences on the particles in space. This problem seems more serious than the above first problem for the $\psi$-ontology. Second, the universal wave function has causal influences on the particles, but not vice versa, why? This is like the back-reaction problem for the $\psi$-ontology. Third, the universal wave function does not play a role that is analogous to that of the classical Hamiltonian; the former is assumed to have causal influences on the particles, but the later does not. 
This means that the usual advantages of the nomological view over the $\psi$-ontology no longer exist for primitivism. 

In fact, primitivism with the above assumption will introduce more problems. 
One problem is that a local measurement on an isolated system may not have a local causal explanation.  
For example, consider a measurement of the energy of a particle in one of its energy eigenstates in a box. The shift of the pointer of the measuring device is not caused by the ontic state of the measured particle in the box, but by the law of nature which does not exist in the box and anywhere in space. 
This new type of nonlocality exists for products states in quantum mechanics or effective product states in Bohmian mechanics. It is beyond the Bell-type nonlocality. Moreover, it is not nonlocality between spacelike separated regions, but nonlocality between one region and nowhere in space (the law of nature). 



Finally, it is worth noting that assuming a law of nature has specific causal influences on a concrete physical entity is inconsistent with the standard distinction between a law of nature and a physical entity as emphasized before.  
The standard view is that physical entities have causal influences upon each other, while a law of nature describes or dictates how physical entities interact with each other. 
For example, a physical system interacts with a measuring device during a measurement, and 
the behaviour of the measuring device such as the probabilities of different results is determined by the ontic state of the system, along with the physical properties of the measuring device. This is one key assumption of the ontological models framework, on which the proof of the $\psi$-ontic view such as the PBR theorem is based. 
In my view, it would be a category mistake to endow a law of nature with the specific causal role of a concrete physical entity.  


\section{A possible way out}

The above analysis suggests that the only sensible way to hold the nomological view of the universal wave function in Bohmian mechanics is to avoid the result of the PBR theorem by rejecting one or more assumptions of the theorem. Let's see if this is possible. 

As we have seen, the PBR theorem is proved based on three preconditions: (1) the quantum algorithm; (2) the ontological models framework; and (3) the preparation independence assumption. 
Bohmian mechanics is consistent with the quantum algorithm. 
Moreover, it also accepts the preparation independence assumption, since two unentangled systems (whose wave function is a product state) have independent ontic states in the theory. 
The crux is whether Bohmian mechanics also accepts the ontological models framework when assuming the nomological view of the universal wave function.\footnote{It has been recently argued that the ontological models framework is wrong in representing quantum theories (Carcassi et al, 2022). My opinion is that this claim is debatable (see also Hubert, 2023). Even if the ontological models framework is wrong, a realist can still prove the result that two orthogonal states correspond to two distinct ontic states (e.g. by an analysis of the Mach-Zehnder interferometer where the same ontic state cannot result in two different definite interference results). This result is enough for us to argue against the nomological view; on this view, two orthogonal states such as two energy eigenstates of a particle in a box may correspond to the same ontic state, i.e. the two Bohmian particles being in these two states may be at rest in the same position in the box. Besides, it is worth noting that the question of whether Bohmian mechanics is consistent with the ontological models framework has been discussed by several authors (Feintzeig, 2014; Leifer, 2014; Drezet, 2015, 2022). Here I will focus on the issue of whether the nomological view of the wave function is consistent with the ontological models framework.} 

The ontological models framework has two fundamental assumptions. 
The first assumption is a realist assumption, which says that if a physical system is prepared such that quantum mechanics assigns a wave function to it, then after preparation the system has a well-defined set of physical properties or an underlying ontic state. 
This assumption is accepted by Bohmian mechanics. According to the nomological view, a subsystem of the universe which has an effective wave function is composed of particles, and the positions of these particles is the ontic state of this subsystem. Note that the ontic state of a subsystem also includes the disposition of its particles which determines their velocities via the guiding equation according to the dispositionalism about laws of nature (Esfeld et al, 2014). 

The second assumption of the ontological models framework says that when a measurement is performed, the behaviour of the measuring device is determined by the ontic state of the measured system, along with the physical properties of the measuring device. 
For a projective measurement $M$, this means that the ontic state $\lambda$ of a physical system determines the probability $p(k|\lambda,M)$ of different results $k$ for the measurement $M$ on the system. 
Then, in order that the nomological view of the universal wave function is valid in Bohmian mechanics, the theory must reject this assumption. 
Concretely speaking, the revised assumption will be that 
when a measurement is performed, the behaviour of the measuring device is determined not only by the (complete) ontic state of the measured system and the measuring device, but also by something else not in the ontology, a nomological component represented by the effective wave function of the system. In particular, for a projective measurement $M$, the ontic state $\lambda$ of a physical system and its effective wave function $\psi$ both determine the probabilities of different results $k$ for the measurement $M$ on the system, which may be denoted by $p(k|\lambda,\psi, M)$.\footnote{Note that when the effective wave function is not nomological but related to the state of reality, the second assumption of the ontological models framework should not be revised this way but keep unchanged, since the complete ontic state $\lambda$ already includes all parts of the state of reality (see also Leifer, 2014; Drezet, 2015).}

It can be seen that the proof of the PBR theorem cannot go through based on this revised assumption. 
Let's remind the basic proof strategy of the PBR theorem. Assume there are $N$ nonorthogonal quantum states $\psi_i$ ($i$=1, ... , $N$), which are compatible with the same ontic state $\lambda$. 
According to the second assumption of the ontological models framework, the ontic state $\lambda$ determines the probability $p(k|\lambda,M)$ of different results $k$ for a measurement $M$. Moreover, there is a normalization relation for any $N$ result measurement: $\sum_{i=1}^{N}p(k_i|\lambda,M)=1$. 
Since there is an $N$ result measurement that satisfies the condition that the first state gives zero Born probability to the first result and the second state gives zero Born probability to the second result and so on, there will be a relation $p(k_i|\lambda,M)=0$ for any $i$, which contradicts the normalization relation. 

Now if the second assumption of the ontological models framework is replaced by the revised assumption, namely that the probabilities of different results $k$ for a measurement $M$ on a physical system is determined not only by the ontic state $\lambda$ of the system, but also  by its wave function  $\psi$, i.e. $p(k|\lambda,M)$ is replaced by $p(k|\lambda,\psi, M)$, then the above contradiction cannot be derived. The reason is as follows. 
Under the revised assumption, the original normalization relation for an $N$ result measurement $\sum_{i=1}^{N}p(k_i|\lambda,M)=1$ holds true only for systems with the same wave function, and for systems with different wave functions $\psi_j$ ($j=1, ..., N$), the normalization relation should be $\sum_{j=1}^{N}\sum_{i=1}^{N}p(k_i|\lambda,\psi_j, M)=1$. 
Then, even if there is an $N$ result measurement that satisfies the condition that the first state gives zero Born probability to the first result and the second state gives zero Born probability to the second result and so on, it will only lead to the relation $p(k_i|\lambda,\psi_i, M)=0$ for any $i$. But this relation does not contradict the new  normalization relation.\footnote{Similarly, the proof that different orthogonal states correspond to different ontic states can also be blocked by the revised assumption.}  

Although the above revised assumption 
can help the nomological view of the universal wave function 
avoid the result of the PBR theorem, it has several issues. 
First of all, the revised assumption already admits that the wave function is real for a single system. 
According to the revised assumption, the probabilities of different results $k$ for a measurement $M$ on a physical system is determined not only by the ontic state $\lambda$ of the system, but also by its effective wave function  $\psi$, i.e. it replaces the response function $p(k|\lambda,M)$ with $p(k|\lambda,\psi, M)$. 
If the wave function is not real for a single system, then the response function for a single system should not explicitly depend on the effective wave function of the system.   
However, it is arguable that the wave function being real for a single system should not be assumed before our analysis; rather, it should be a possible result obtained at the end of our analysis. For it directly excludes the possibility that the wave function is not real for a single system (e.g. the $\psi$-epistemic view). This is unsatisfactory. 

Next, and more seriously for a realist view, 
the results of measurements and their probabilities cannot be explained in ontology under the revised assumption. 
According to the original assumption of the ontological models framework, when a measurement is performed, the behaviour of the measuring device is fully determined by the complete ontic state of the measured system and the measuring device. 
But according to the revised assumption, the behaviour of the measuring device is not fully determined by the complete ontic state of the measured system and the measuring device. 
Then, the results of measurements and their probabilities will be unexplainable in ontology. 

Let me give a simple example. 
Consider a measurement of an observable (other than position) on a system being in an eigenstate of the observable such as a measurement of the energy of a particle in one of its energy eigenstates in a box. 
The measurement will always yield a definite result, the corresponding energy eigenvalue. 
According to the ontic view of the effective wave function, this measurement result is determined by the ontic state of the measured system represented by its wave function, and thus it is explainable in ontology. 
This is consistent with the original assumption of the ontological models framework. 
But according to the nomological view of the universal wave function, 
the ontic state of the measured system is only the position of its particle which is at rest in the box, and it does not include the energy eigenstate of the system. 
Since the particle may be in the same position in the box for different energy eigenstates, the ontic state of the measured system cannot determine the measurement result, which are different for different energy eigenstates. 
This is just what the revised assumption says. 
Thus, when assuming the nomological view of the universal wave function or according to the revised assumption, the definite results of certain measurements will be unexplainable in ontology. 

Finally, it seems that one can even argue that the revised assumption cannot be true. 
Consider two measurement situations in each of which there are a measured system and a measuring device. The two measured systems are identical, so do the two measuring devices. Moreover, 
the initial ontic states of the two measured systems and measuring devices are the same, but the effective wave functions of the two measured systems are different.\footnote{This is possible as can be seen from the previous example. In the example, the effective wave function of the measured system is an energy eigenstate in a box. The ontic state of the measured system, which is represented by the position of its particle in the box, may be the same for different energy eigenstates. Moreover, the ontic states of the measuring devices are supposed to be the same before different measurements.}   
Since these two situations cannot be distinguished in ontology, the laws of motion for them must be the same. 
Then, the behaviours of the measuring devices will be the same for the two measurements, and thus they cannot depend on the effective wave functions of the measured systems, which are different for the two measurements. 
This means that the revised assumption cannot be true.\footnote{There is also a similar example which can reveal the problem of the nomological view without referring to measurements. Suppose there are two electrons which have different effective wave functions but whose Bohmian particles have the same position and velocity initially. Then, Bohmian mechanics will predict that these two  Bohmian particles may have different positions and velocities at a later time. Since no laws of motion can lead to that two identical particles, which have the same ontic state initially, have different ontic states later, it seems that there cannot be only particles in ontolgy in Bohmian mechanics. I will analyze this issue in more detail in future work.} 


\section{What is the $\psi$-ontology?}

I have argued that the PBR theorem as a $\psi$-ontology theorem requires that 
the ontology of Bohmian mechanics cannot consist only in particles, and it must include the $\psi$-ontology. The next question is: what is the $\psi$-ontology or the ontology represented by the wave function? 
For now, there are three $\psi$-ontologies, namely a physical field in a fundamental high-dimensional space as in wave function realism (Albert, 1996, 2013; Ney, 2021), a multi-field in three-dimensional space (Forrest, 1988; Belot, 2012; Hubert and Romano, 2018; Romano, 2021), and RDMP (Random Discontinuous Motion of Particles) in three-dimensional space (Gao, 1993, 2017, 2020, 2022).\footnote{According to the RDMP interpretation of the wave function, a quantum system is composed of particles with mass and charge, which undergo random discontinuous motion in three-dimensional space, and the wave function represents the propensities of these particles which determine their motion, and as a result, the state of motion of particles is also described by the wave function. At each instant all particles have a definite position, while during an infinitesimal time interval around each instant they move throughout the whole space where the wave function is nonzero in a random and discontinuous way, and the probability density that they appear in every possible group of positions in space is given by the modulus squared of the wave function there.} 
In this section, I will briefly discuss them. 

As noted before, 
there are two major objections to the $\psi$-ontology raised by the proponents of the nomological view of the universal wave function (D\"{u}rr, Goldstein and Zangh\`{i}, 1997; Goldstein and Zangh\`{i}, 2013). 
The first one is that the wave function of a many-particle system is defined not in three-dimensional space, but in the high-dimensional configuration space, the set of all hypothetical configurations of the system. Thus it seems untenable to view the wave function as directly describing a real physical field. 
This objection applies to wave function realism,\footnote{The proponents of wave function realism may  insist that the high-dimensional configuration space is a real, fundamental space, while the three-dimensional space we perceive is somehow illusory. However, this leads to the well-known problem of how to explain our three-dimensional impressions (see Ney, 2021, chap.8 for the latest analysis of this problem).} but it does not apply to the multi-field ontology and the RDMP ontology, which exist in three-dimensional space. 

The second objection is that although the wave function affects the behavior of the configuration of the particles according to the guiding equation, there is no back action of the configuration upon the wave function according to the Schr\"{o}dinger equation. Then, since a physical entity is supposed to satisfy the action-reaction principle, the wave function cannot describe a physical entity in Bohmian mechanics. 
This objection applies to wave function realism and the multi-field ontology, 
but it does not apply to the RDMP ontology, for which the wave function is not a physical entity different from particles but a property of the particles that undergo random discontinuous motion.\footnote{It seems that even for wave function realists and multi-field ontologists this objection is not a very strong argument either. They may claim that the physical entity represented by the wave function is a new type of entity and it is not subject to back reaction (see, e.g. Hubert and Romano, 2018; Romano, 2021).} 

Besides being able to answer these objections, the RDMP ontology may also avoid the potential problems of Bohmian mechanics. First, the guiding equation is not unique and it is also ill-defined at the nodes of the wave function in Bohmian mechanics.\footnote{However, global existence and uniqueness for the Bohmian dynamics can be proven (Berndl et al, 1995; Teufel and Tumulka, 2005).} 
For the RDMP ontology, 
the guiding equation is replaced by the random dynamics $P(Q(t), t) = |\Psi(Q(t), t)|^2$, which means that at every instant $t$ the particle configuration of the universe is definite but random, and its probability of being a given $Q(t)$ is equal to the Born probability $|\Psi(Q(t), t)|^2$. The random dynamics is unique and has no ill-definedness. In some sense, we may say that the particles are guided by the wave function in a probabilistic way for the RDMP ontology. 

Second, the status and justification of the quantum equilibrium hypothesis, which is needed for accounting for the empirical equivalence between Bohmian mechanics and quantum mechanics, remains a controversial issue (see Callender, 2007 and Norsen, 2018 for helpful review). For the RDMP ontology, no additional quantum equilibrium hypothesis is needed. 
No matter what the initial probability distribution of the configuration of the particles is, the latter probability distribution will (typically) be consistent with the Born distribution according to the random dynamics.\footnote{Certainly, this consistency is still obtained by the postulate that the RDMP dynamics is in agreement with Born's distribution. But it is arguable that one unified postulate is better than two separate postulates. Moreover, there are plausible arguments supporting the RDMP ontology (see Gao, 2017, 2020, 2022).}   
In this sense, we may say that the random dynamics unifies the guiding equation and the quantum equilibrium hypothesis.

The unification also makes the RDMP ontology avoid the problem of the double role of the wave function. 
In Bohmian mechanics, the wave function has a double role. On the one hand, the wave function determines the velocities of the Bohmian particles via the guiding equation. On the other hand, the wave function also determines the initial probability distribution of the configuration of the particles via the quantum equilibrium hypothesis. 
Why the wave function plays this double role and whether the two roles are related is still an unresolved issue of Bohmian mechanics. 
For the RDMP ontology, the wave function has a unique role, namely determining the random discontinuous motion of particles via the random dynamics. 

Third, by the same reasoning as for the $\psi$-ontology, 
it can be argued that mass and charge should be also included in the ontology of quantum mechanics (Gao, 2022a). 
For the RDMP ontology, 
the particles have mass and charge and they also interact with each other. 
By comparison, the Bohmian particles are usually assumed to have no mass and charge (on the nomological view) and they have no interaction with each other (when their effective wave function is a product state).\footnote{Even if the Bohmian particles are assumed to have mass and charge, these properties have no the causal efficacy required in ontology such as deviating the pointer of a measuring device. By contrast, the mass and charge of a particle in the RDMP ontology have such required causal efficacy.} 
In addition, mass and charge are not included in the ontologies of wave function realism and the multi-field interpretation, and it seems that they can hardly be included in these field ontologies (Gao, 2022a). 

Finally, it should be noted that the problem of ontology, such as what the $\psi$-ontology is, is independent of how to solve the measurement problem. 
The former refers only to the ontology, while the latter is also related to the dynamics for the ontology. 
It seems that the RDMP ontology alone, unlike the continuous motion of particles in Bohmian mechanics, cannot solve the measurement problem. 
However, it has been argued that the RDMP ontology fits well with collapse theories, another promising solution to the measurement problem, since the random motion of particles can provide a natural collapse noise for these theories (Gao, 2017). 
Admittedly, how to solve the measurement problem with the RDMP ontology is still an open issue, and it deserves further study (for a recent attempt see Gao, 2022b).  


\section{Conclusions}

It is widely thought that the nomological view of the wave function in Bohmian mechanics, which says that the universal wave function is not ontic but nomological, is consistent with the PBR theorem, which proves that the effective wave function in Bohmian mechanics is ontic. 
In this paper, I argue that this received view is debatable. 
First, the nomological view of the universal wave function is not compatible with the ontic view of the effective wave function. This incompatibility result holds true not only for Humeanism and dispositionalism but also for primitivism about laws of nature. 
Next, although the nomological view can be held by rejecting one key assumption of the ontological models framework on which the PBR theorem is based, the rejection will lead to serious problems, such as that the results of measurements and their probabilities cannot be explained in ontology in Bohmian mechanics. 
This new analysis suggests that the ontology of Bohmian mechanics consist in both particles and the wave function.  
Finally, I briefly discuss three $\psi$-ontologies and argue that the RDMP (Random Discontinuous Motion of Particles) ontology can answer the objections to the $\psi$-ontology raised by the proponents of the nomological view. 
  

\section*{References}
\renewcommand{\theenumi}{\arabic{enumi}}
\renewcommand{\labelenumi}{[\theenumi]}
\begin{enumerate}



\item{} Albert, D. Z. (1996). Elementary Quantum Metaphysics. In J. Cushing, A. Fine and S. Goldstein (eds.), Bohmian Mechanics and Quantum Theory: An Appraisal. Dordrecht: Kluwer, 277-284.

\item{} Albert, D. Z. (2013). Wave function realism. In Ney, A. and D. Z. Albert (eds.),  The Wave Function: Essays on the Metaphysics of Quantum Mechanics. Oxford: Oxford University Press. pp. 52-57. 

\item{} Allori, V., S. Goldstein, R. Tumulka, and N. Zangh\`{i} (2008). On the common structure of Bohmian mechanics and the Ghirardi-Rimini-Weber theory, British Journal for the Philosophy of Science 59 (3), 353-389.


\item{} Belot, G. (2012). Quantum states for primitive ontologists: A case study
European Journal for Philosophy of Science, 2(1): 67-83.

\item{} Berndl, K., D\"{u}rr, D., Goldstein, S., Peruzzi, G. and Zangh\`{i}, N. (1995). On the Global Existence of Bohmian Mechanics. Communications in Mathematical Physics, 173, 647-673.

\item{} Bohm, D. (1952). A suggested interpretation of quantum theory in terms of ``hidden" variables, I and II. Physical Review 85, 166-193.

\item{} Callender, C. (2015). One world, one beable. Synthese, 192, 3153-3177.

\item{}  Carcassi, G. Oldofredi, A. and Aidala, C. A. (2022). On the reality of the quantum state once again: A no-go theorem for $\psi$-ontic models. arXiv:2201.11842v2 [quant-ph]. 

\item   Colbeck, R. and Renner, R. (2012). Is a system's wave function in one-to-one correspondence with its elements of reality? Physical Review Letters, 108, 150402.

\item   Colbeck R. and Renner R. (2017). A system's wave function is uniquely determined by its underlying physical state. New J. Phys. 19, 013016.

\item  de Broglie, L. (1928). La nouvelle dynamique des quanta. In: Electrons et Photons: Rapports et Discussions du Cinquieme Conseil de Physiquepp, pp. 105–132. Gauthier-Villars, Paris. English translation: Bacciagaluppi, G., Valentini, A.: Quantum Theory at the Crossroads: Reconsidering the 1927 Solvay Conference. Cambridge University Press, Cambridge (2009). 

\item{} Dorato, M. (2015) Laws of nature and the reality of the wave function. Synthese, 192, 3179-3201.

\item{} Dorato, M. and M. Esfeld (2015). The metaphysics of laws: dispositionalism vs. primitivism. In: Tomasz Bigaj and Christian Wüthrich (eds.), Metaphysics in Contemporary Physics Poznań Studies in the Philosophy of the Sciences and the Humanities, vol. 104), pp. 403-424. Amsterdam/New York, NY: Rodopi, Brill.

\item{} Drezet, A. (2015). The PBR theorem seen from the eyes of a Bohmian. International Journal of Quantum Foundations 1, 25-43.

\item{}  Drezet, A. (2022). A no-go theorem for psi-anomic models under the restricted ontic indifference assumption. International Journal of Quantum Foundations 8, 16-30.

\bibitem{} D\"{u}rr, D., S. Goldstein,  and N. Zangh\`{i} (1992). Quantum equilibrium and the origin of absolute uncertainty. Journal of Statistical Physics 67, 843-907.

\bibitem{} D\"{u}rr, D., Goldstein, S. and  Zangh\`{i}, N.  (1997). Bohmian mechanics and the meaning of the wave function. In R. S. Cohen,  M. Horne, and J. Stachel (eds.), Experimental Metaphysics - Quantum Mechanical Studies for Abner Shimony, Volume One; Boston Studies in the Philosophy of Science 193. Dordrecht: Kluwer. pp. 25-38. 

\item Esfeld, M., Lazarovici, D., Hubert, M. and D\"{u}rr, D. (2014). The ontology of Bohmian mechanics. British Journal for the Philosophy of Science.  65 (4), 773-796.


\item{} Feintzeig, B. (2014). Can the ontological models framework accommodate Bohmian mechanics? Studies in History and Philosophy of Modern Physics, 48, 59-67.

\item{} Forrest, P. (1988). Quantum Metaphysics. Oxford: Basil Blackwell.

\bibitem{} Gao, S. (1993). A suggested interpretation of quantum mechanics in terms of discontinuous motion (unpublished manuscript).

\item Gao, S. (2017). The Meaning of the Wave Function: In Search of the Ontology of Quantum Mechanics. Cambridge: Cambridge University Press.

\item Gao, S. (2020). A Puzzle for the Field Ontologists. Foundations of Physics 50, 1541-1553. 


\item Gao, S. (2022a). Reality of mass and charge and its implications for the meaning of the wave function. http://philsci-archive.pitt.edu/21317/.

\item Gao, S. (2022b). On Bell’s Everett (?) Theory. Foundations of Physics 52, 89. 

\item{} Goldstein, S. (2021). Bohmian Mechanics, The Stanford Encyclopedia of Philosophy (Fall 2021 Edition), Edward N. Zalta (ed.), https://plato.\\stanford.edu/archives/fall2021/entries/qm-bohm/.

\bibitem{} Goldstein, S., and S. Teufel (2001). Quantum spacetime without observers: Ontological clarity and the conceptual foundations of quantum gravity. In Callender, C. and Huggett, N., eds., Physics meets Philosophy at the Planck Scale, Cambridge: Cambridge University Press. pp.275-289.

\bibitem{}  Goldstein, S.  and Zangh\`{i}, N. (2013). Reality and the role of the wave function in quantum theory. In Ney, A. and D. Z. Albert (eds.). The Wave Function: Essays on the Metaphysics of Quantum Mechanics. Oxford: Oxford University Press.  pp. 91-109.

\item{} Hardy, L. (2013). Are quantum states real? International Journal of Modern Physics B 27, 1345012.

\item{} Harrigan, N. and Spekkens, R. (2010). Einstein, incompleteness, and the epistemic view of quantum states. Foundations of Physics 40, 125-157.

\item{} Hubert, M. (2023). Is the Statistical Interpretation of Quantum Mechanics $\psi$‑Ontic or $\psi$‑Epistemic? Foundations of Physics 53, 16.

\item{} Hubert, M. and Romano, D. (2018). The wave-function as a multi-field. European Journal for Philosophy of Science 8, 521-537. 
  
 \item{} Leifer, M. S. (2014). Is the quantum state real? An extended review of $\psi$-ontology theorems, Quanta 3, 67-155.
 
\item{} Maudlin, T. (2007). The Metaphysics within Physics. Oxford: Oxford University Press. 


  \item   Ney, A. (2021). The World in the Wave Function: A Metaphysics for Quantum Physics. Oxford: Oxford University Press.
  
  \item  Ney, A. and D. Z. Albert (eds.) (2013). The Wave Function: Essays on the Metaphysics of Quantum Mechanics. Oxford: Oxford University Press.


\item{} Oldofredi, A. and Calosi, C. (2021). Relational Quantum Mechanics and the PBR Theorem: A Peaceful Coexistence. Foundations of Physics 51, 82.

\item{}  Oldofredi, A. and L\'{o}pez, C. (2020). On the Classification Between $\psi$-Ontic and $\psi$-Epistemic Ontological Models. Foundations of Physics 50, 1315-1345. 

 \item Pusey, M., Barrett, J. and Rudolph, T. (2012). On the reality of the quantum state. Nature Phys. 8, 475-478. 

 \item  Romano, D. (2021). Multi-field and Bohm's theory. Synthese 198, 10587-10609. 

 \item Rovelli, C. (1996). Relational quantum mechanics. Int. J. Theor. Phys. 35(8), 1637-1678. 

 
 \item  Teufel, S. and Tumulka, R. (2005). Simple Proof for Global Existence of Bohmian Trajectories. Communications in Mathematical Physics, 258, 349-365.
 
  \item  Valentini, A. (2020). Foundations of statistical mechanics and the status of the Born rule in de Broglie-Bohm pilot-wave theory. In V. Allori (ed.), Statistical Mechanics and Scientific Explanation: Determinism, Indeterminism and Laws of Nature. World Scientific. pp. 423-477.


\end{enumerate}

 
\end{document}